\documentclass[aip,apl,preprint,floatfix,showpacs]{revtex4-1}
\usepackage{graphicx}% Include figure files
\usepackage[latin1]{inputenc}
\usepackage[bf,SL,BF]{subfigure}

\newcommand{\ket}[1]{\left| #1 \right>}

\newcommand{\be}{ \begin{equation} }
\newcommand{\ee}{ \end{equation} }
\newcommand{\bea}{\begin{eqnarray} }
\newcommand{\eea}{\end{eqnarray} }

\begin{document}

\title{Quantum interference and control of the optical response in quantum dot molecules}
\author{H. S. Borges, L. Sanz, J. M. Villas-Boas and A. M. Alcalde}
\affiliation{Instituto de Física, Universidade Federal de Uberlândia, 38400-902, Uberlândia-MG, Brazil}

\begin{abstract}
We discuss the optical response of a quantum molecule under the action of two lasers fields. Using a realistic model and parameters, we map the physical conditions to find three different phenomena reported in the literature: the tunneling induced transparency, the formation of Autler-Townes doublets, and the creation of a Mollow-like triplet. We found that the electron tunneling between quantum dots is the responsible for the different optical regime. Our results not only explain the experimental results in the literature but also give insights for future experiments and applications in optics using quantum dots molecules.
\end{abstract}
\pacs{73.21.La, 42.50.-p, 73.40.Gk, 03.65.Yz} \keywords{Quantum
dots molecule, excitons, quantum interference.}
\maketitle

Quantum interference and coherent effects have been widely studied
and experimentally reported in various schemes of three-level
systems ($\Lambda-$, $V-$ and cascade- type).
Driven by two coherent optical fields,
atomic systems and single charged quantum dots have been used to
investigate quantum phenomena such as: Autler-Townes splitting
(ATS)~\cite{Yang05}, Mollow triplets~\cite{Mollow69,Ulhaq12} and
electromagnetically induced transparency (EIT)~\cite{Marangos05}.
The optical response of a physical system is considerably enriched when
a multi-level structure is taking into account~\cite{Lu98,Ottaviani06}.
A quantum dot molecule (QDM), formed by coherent tunneling between two individual semiconductor quantum dots, is an ideal system to investigate quantum interference processes similar to those reported in atomic systems~\cite{Katz95,Wang04}, as it presents a controllable multilevel excitonic structure~\cite{Doty08}.

In a previous work, we analyze the behavior of tunneling induced transparency (TIT) and slow light
effects in a QDM modeled by a $\Lambda$-type system, where the role of the optical control field is replaced by tunneling coupling
parameter between quantum dots~\cite{Borges12}. Here, we investigate the optical susceptibility of a QDM coupled by tunneling, coherently driven
by a probe and a control lasers. We demonstrate that the laser fields creates a level configuration that can be described as two optically active three-level subsystems where tunneling induced transparency (TIT) and ATS can be achieved. We also show the physical conditions for the formation of a Mollow-like triplet, which depends on the ratio between the control laser coupling and the tunneling.

To model the system, we considered a QDM composed of two vertically aligned QD of different sizes separated by a barrier of width $d$. The barrier width as well as the structural asymmetry of the system inhibits the tunneling of holes~\cite{Bracker06} and we will neglect it here. We also considered the weak probe laser with frequency $\omega_p$ nearly in resonance with direct exciton transition of the ``top" dot, while we tuned the control laser frequency $\omega_c$ into resonance with direct exciton transition of the ``bottom" dot. Additionally, we use an electric field $F$ applied along the growth direction to control the QDM level alignment, and include the effects of radiative spontaneous decay and pure dephasing of all excitonic states, described here by rates $\Gamma^j_0$ and $\gamma_j$, respectively.

\begin{figure}[tb]
\includegraphics[scale=0.65]{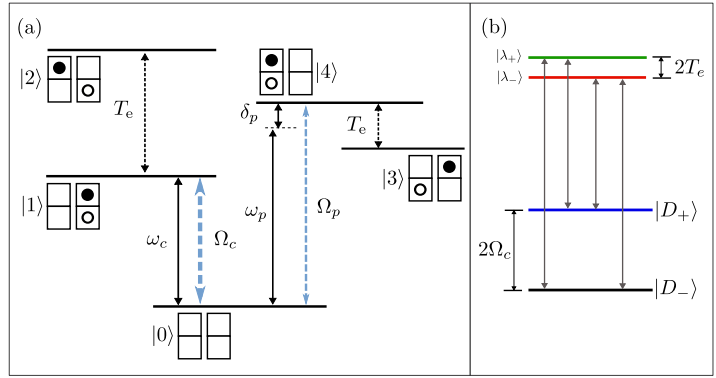}
\caption{(Color online) (a) Schematic representation of the Hamiltonian (\ref{eq:htrans}) in the bare state basis including the optical and tunneling couplings. (b) Schematic representation of dresses states at $F_+$ for the condition $\Omega_p\ll\Omega_c$, gray lines represent the allowed optical transitions.}
\label{fig:levels}
\end{figure}

Considering the above assumption and, in the limit of low excitation, we  can describe the system with a five-states basis where $\ket{0}$ describes the exciton vacuum, and $\ket{1}$ ($\ket{4}$) the direct exciton in the top (bottom) QD. The indirect exciton state $\ket{2}$ ($\ket{3}$) is obtained by one electron tunneling from the top (bottom) to the bottom (top) QD.
Under electric-dipole and rotating-wave approximations and after performing the unitary transformation $U=e^{-i\omega_c t}(|1\rangle \langle 1|+|2\rangle \langle 2|)+e^{-i\omega_p t}(|3\rangle \langle 3|+|4\rangle\langle 4|)$, which removes the time-dependent oscillatory terms, the matrix form of Hamiltonian %(\ref{eq:hor})
is given by:
\begin{equation}
H^{'}=\left(\begin{array}{ccccc}0&\hbar\Omega_c&0&0&\hbar \Omega_p\\
\hbar\Omega_c&0&T_e&0&0\\0&T_e&\delta_2+\Delta_F&0&0\\
0&0&0&\delta_3-\Delta_F&T_e\\\hbar \Omega_p&0&0&T_e&\delta_p
\end{array}\right),
\label{eq:htrans}
\end{equation}
where $T_e$ is the electron tunneling matrix element, $\delta_p=\hbar\left(\omega_{40}-\omega_p\right)$ is the probe laser detuning between vacuum and bottom QD states, $\delta_2=\hbar\omega_{21}$ is the direct-indirect exciton detuning in the top dot, $\delta_3= \delta_p+\hbar\omega_{34}$ and $\Delta_F=eFd$ is the energy shift due to the gate field. The Rabi frequencies associated with control and probe laser field are $\Omega_{c}={\mu_{01}\mathcal{E}_{c}}/{2\hbar}$ and $\Omega_{p}={\mu_{04}\mathcal{E}_{p}}/{2\hbar}$ respectively, where $\mu_{ij}$ is the dipole momentum matrix element from state $i$ to $j$ and $\mathcal{E}$ is the laser electric field amplitude. Here we have used $\omega_{ij}=(E_{i}-E_{j})$, where $E_{i}$ is the energy of state $\ket{i}$. A schematic representation of the Hamiltonian for a  fixed  electric field $F$ is shown in Fig.~\ref{fig:levels}(a).

For our simulations we used realistic parameters for InAs self-assembled~\cite{Bracker06}, but our conclusions can be applied to other QDM made of different materials. In this system, the typical effective decay rate is of the order of $\Gamma_j=\Gamma_{0}^{j}/2+\gamma_j\sim 2-10~\mu \mathrm{eV}$~\cite{Borri03, Bardot05} for direct exciton states, with indirect excitons rates being three times smaller ~\cite{Butov94}. The realistic bare exciton energies are the same by Rolon and Ulloa~\cite{Rolon10}. The tunneling coupling varies from $0.01$ to $0.1\mathrm{meV}$~\cite{Nakata00} and $1$ to $10~\mathrm{meV}$~\cite{Emay07}, for weak and strong tunneling regime respectively. The parameters associated to the susceptibility such as the optical confinement factor $\Gamma_{opt}=6\times 10^{-3}$, momentum matrix element $\mu_{40}/e=21\sim\mathrm{\AA}$, and volume $V$, were taken from Kim \textit{et al.}~\cite{Kim04}. Specifically, we set $\Gamma_4=10~\mu\mathrm{eV}$, $\Omega_p=0.25\Gamma_4$ and $\Omega_c=5\Gamma_4=0.05$ meV, following the condition $\Omega_p \ll \Omega_c,T_e$.

In order to better understand the optical response of the QDM, we compute the eigenvalues of the Hamiltonian (\ref{eq:htrans}) as a function of external electric field $F$. This shows several anticrossings. For instance, around $F\sim 0$ is observed an anticrossing between the two indirect exciton (not shown here). Nevertheless, we focus our interest on the electric field region $F_\pm$ where direct and indirect excitons anticrosses. Around this region the indirect exciton can be efficiently populated~\cite{Borges10}, which is the key ingredient for TIT effect~\cite{Borges12}. A more detailed analysis shows that the electric field condition $\Delta_{F_+} = \omega_{34}$ and $\Delta_{F_-} = -\omega_{21}$ guarantees a large population of indirect exciton states $\ket{3}$ and $\ket{2}$ respectively. We choose to analyze the optical response for positive values of electric field. In this situation, the condition $\Delta_{F_+} = \omega_{34}$ provides $F_+ = 22.7~\mathrm{kV/cm}$ where we expect the action of interference effects on the optical properties.

\begin{figure}[tb]
\includegraphics[scale=0.6]{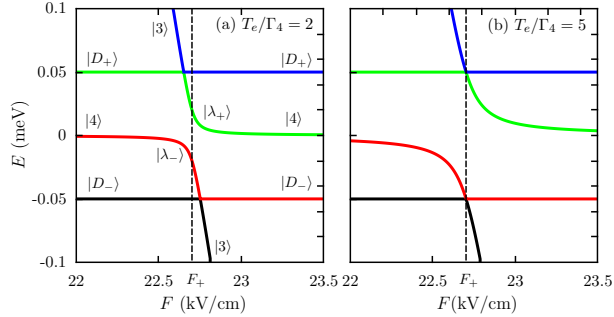}
\caption{(Color online) Exciton dressed energies around $F_+= 22.7~\mathrm{kV/cm}$ (dashed vertical line) for (a) $T_e/\Gamma_4 = 2$ and (b) $T_e/\Gamma_4 = 5$. Note that energy level structure is sensitive to the action of electron tunneling. In all cases $\Omega_c = 5 \Gamma_4$ is considered. Labels on (a) correspond to the eigenstates for electric field values far from $F_+$.}
\label{fig:espectro}
\end{figure}

At $F_+$, the strong coupling field $\Omega_c$ creates two superpositions of exciton vacuum and top direct exciton, $\ket{D_\pm}=(\ket{0}\pm\ket{1})/\sqrt{2}$, which are energetically separated by $2\Omega_c$. Since the probe laser $\Omega_p$ is weak ($\Omega_p\ll\Omega_c$), it will not create a dressed state as in the case of the strong coupling field $\Omega_c$. On the other hand, the action of tunneling around $F_+$ creates the states $\ket{\lambda_{\pm}}$ which are $2T_e$ apart. In the weak tunneling regime ($T_e\ll\Omega_c$) and away from $F_+$, the states are approximately $\ket{\lambda_-} \approx \ket{3}$ and $\ket{\lambda_+} \approx \ket{4}$, otherwise the states $\ket{\lambda_{\pm}}$ become superpositions of bare states $\ket{3}$ and $\ket{4}$. A schematic representation of the dressed states with energies splitting and allowed optical transitions are shown in Fig.~\ref{fig:levels}(b). At $E=\Omega_c$($-\Omega_c$), the state $\ket{D_+}$($\ket{D_-}$) anticross with $\ket{\lambda_-}$ ($\ket{\lambda_+}$) as can be seen in Fig.~\ref{fig:espectro}, where we plot the energy spectrum around $F_+$ for (a) $T_e/\Gamma_4=2$ and (b) $T_e/\Gamma_4=5$. The optical response of the QDM can be described as the combined action of two subsystems: I) a three level system composed by $\left\{\ket{D_+},\ket{\lambda_-},\ket{\lambda_+}\right\}$ optically relevant at $\delta_p \sim \Omega_c$ and II) a three level system composed by $\left\{\ket{D_-},\ket{\lambda_-},\ket{\lambda_+}\right\}$ optically active at $\delta_p \sim -\Omega_c$. A simple comparison of Figs.~\ref{fig:espectro}(a) and (b) shows that the dressed excitonic spectra is strongly modified by tunneling. Thus, it is expect that the tunneling coupling modifies the interplay between subsystems I) and II), as we discuss below.

To obtain the dynamics and optical properties of the QDM, we numerically solve the Liouville-von Neumman-Lindblad equation, as in our previous works~\cite{Borges10,Borges12}. In order to analyze the optical properties associated with the probe field, we evaluate the optical susceptibility,
 $\chi=\frac{\Gamma_{opt}}{V}\frac{|\mu_{40}|^2}{\epsilon_0 \hbar \Omega_p}\rho_{04}$, where $\Gamma_{opt}$ is the optical confinement factor, $V$ is the volume of a single QD, and $\epsilon_0$ is the dielectric constant. The susceptibility is a complex function, written as $\chi=\chi^{\prime}+i\chi{\prime \prime}$, where the absorption coefficient $\alpha(\omega_p)$ is given by $\chi{\prime \prime}$, while the refractive index $n(\omega_p)$ of the probe field depends on both, real and imaginary parts of $\chi$.

In the absence of the control laser ($\Omega_c=0$), the QDM becomes a three-level lambda system involving the vacuum state $\ket{0}$, the direct exciton $\ket{4}$ and the long lived indirect exciton state $\ket{3}$. The absorption coefficient as a function of laser detuning, $\delta_p$, and tunneling, $T_e$, shows a V-like form, centered at $\delta_p=0$~\cite{Borges12}. For $T_e/\Gamma_4<1/2$, the absorption has a transparency window when the probe laser is resonant with the direct exciton transition. This effect, denoted as TIT, is associated with a destructive interference between paths mediated by probe field and tunneling coupling. If $T_e/\Gamma_4>1/2$, it is obtained a Autler-Townes doublet.

\begin{figure}[tb]
\centering
\includegraphics[scale=0.7]{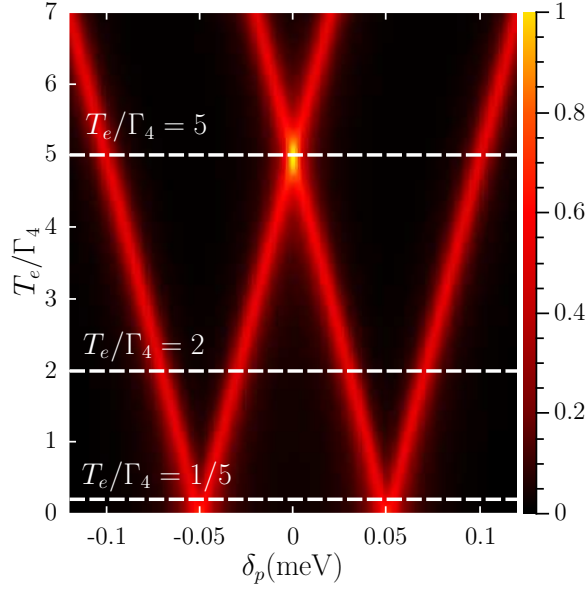}
\caption{(Color online) Imaginary part of optical susceptibility as a function of probe laser detuning $\delta_p$ and the rate $T_e/\Gamma_4$ for $\Omega_c/\Gamma_4=5$. The white dashed lines identifies different optical response regimes of the QDM. Notice the formation of the Mollow-like triplet at $T_e/\Gamma_4=\Omega_c/\Gamma_4=5$.}
\label{fig:tunneling}
\end{figure}

\begin{figure}[ht]
\centering
\includegraphics[scale=0.7]{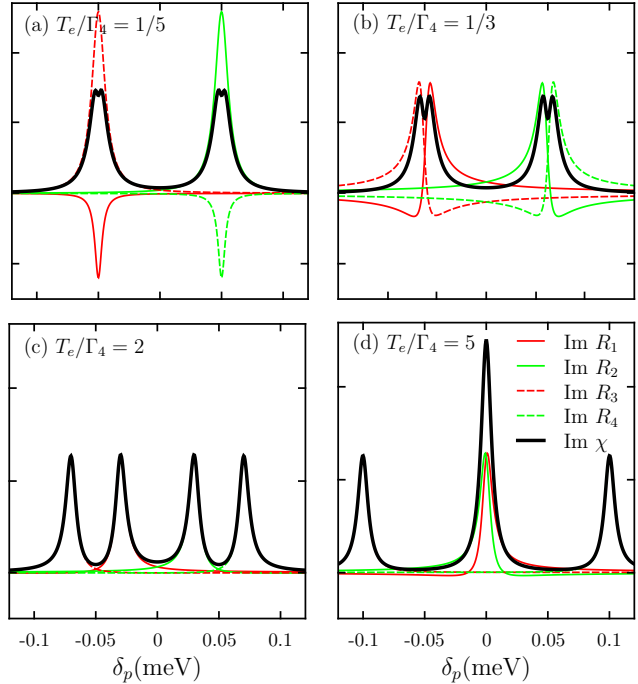}
\caption{(Color online) Total absorption (black solid line) and the imaginary part of $R_i$ (red and green solid and dashed lines) as a function of the probe detuning $\delta_p$ for
$T_e/\Gamma_4$ values corresponding to the white lines in Fig.~\ref{fig:tunneling}:
(a) $T_e/\Gamma_4=1/5$, (b) $T_e/\Gamma_4=1/3$, (c) $T_e/\Gamma_4=2$ and (d)$T_e/\Gamma_4=5$. In all cases, $\Omega_c/\Gamma_4=5$ is considered.}
\label{fig:regimes}
\end{figure}

The effect of $\Omega_c$ is shown in Fig.~\ref{fig:tunneling}, where we plot the imaginary part of susceptibility, as a function of the probe laser detuning $\delta_p$ and the ratio $T_e/\Gamma_4$. The apparition of two V-like branches on the susceptibility, centered at $\delta_p=\pm\Omega_c$, is the signature of the presence of the two distinguishable three-level subsystems mentioned above. Depending on the ratio $T_e/\Gamma_4$, the susceptibility shows three different behaviors. For small values of $T_e/\Gamma_4$, weak tunneling condition, it is observed a TIT-like effect. When $T_e/\Gamma_4$ increases, the optical response features a ATS behavior. It is interesting to notice that, at the exact condition $T_e=\Omega_c$ (fulfilled as $T_e/\Gamma_4=5$ in Fig.~\ref{fig:regimes}), two of the doublets from different branches merge at $\delta_p=0$, with the subsequent creation of a Mollow-like triplet (MT). For $T_e/\Gamma_4>5$, we recover a four peak absorption profile but with the exchange of the central peaks.

In order to establish specific conditions to distinguish the optical behavior as function of $T_e/\Gamma_4$, we extend the procedure used in our previous work~\cite{Borges12}. In the limit of low excitation ($T_e>>\Omega_p$), the density matrix element $\rho_{04}$ can be written as a sum of four components $R_i$:
\begin{equation}
\rho_{04}=\sum^{4}_{i=1}R_i\propto\sum^4_{i=1} \frac{1}{\delta_p-\delta_p^{i}},
\label{eq:ressonances}
\end{equation}
where $\delta_p^1=Z_1-\sqrt{Z_2-2\sqrt{Z_3}}/4$, $\delta_p^2=Z_1+\sqrt{Z_2-2\sqrt{Z_3}}/4$, $\delta_p^3=Z_1-\sqrt{Z_2+2\sqrt{Z_3}}/4$, and $\delta_p^4=Z_1+\sqrt{Z_2+2\sqrt{Z_3}}/4$ with auxiliary parameters given by $Z_1=-i(\Gamma_1+\Gamma_3+\Gamma_4)/4$, $Z_2=16(T_e^2+\Omega_c^2)-(\Gamma_4-\Gamma_3)^2-\Gamma^2_1$ and $Z_3=(\Gamma^2_1-16\Omega_c^2)\left[(\Gamma_4-\Gamma_3)^2-16T_e^2\right]$.

In Fig.~\ref{fig:regimes}, we plot the imaginary part of the susceptibility together (solid lines) with the imaginary part $R_i$ for $T_e/\Gamma_4$ corresponding to the white dashed lines in Fig.~\ref{fig:tunneling}. The absorption profile (solid line) is the result of the contribution of four Lorentzian-like peaks, each corresponding to the four resonances.
For weak coupling, considering $T_e/\Gamma_4<1/4$, the absorption coefficient results from the contribution of two pairs of resonances, each pair with a positive and a negative amplitude, as shown Fig.~\ref{fig:regimes}(a). % for $T_e/\Gamma_4=1/5$.
The presence of negative resonances is a characteristic of destructive quantum interference, which is associated with optical transparency~\cite{Salloum07,Salloum10}. The creation of new interference path reduces the contribution of the negative resonances producing a dip in the absorption spectrum at $\delta_p=\pm\Omega_c$, instead of a transparency window.
In this way, it is convenient to describe this behavior as a \emph{near} TIT regime.

For intermediate coupling, $1/4<T_e/\Gamma_4<1$, we verify that the dip in absorption line increase as a consequence of a gap between the resonances, as is shown in Fig.~\ref{fig:regimes}(b). Notice that the resonances still have negative amplitudes revealing the influence of quantum interference in the transition between near TIT and ATS regimes.
An interesting feature is the reduction of the light group velocity $v_g$, the rate $c/v_g$ varies between $1\times 10^4$ to $5\times 10^4$ for $1/4<T_e/\Gamma_4<1$, with a maxima around $T_e/\Gamma_4\cong0.55$. In contrast, for the weak coupling regime we found no significant reduction of light group velocity.

A full ATS behavior is observed in the strong coupling regime, $T_e/\Gamma_4 > 1$, shown in Fig.~\ref{fig:regimes}(c), where the absorption of probe optical coupling is a result of the contributions of four positive resonance. This behavior is characteristic of the ATS effect and the dips at $\delta_p\simeq \pm\Omega_c$ observed in the absorption spectrum reveals the signature of the electron interdot tunneling.
In the particular case when $\Omega_c=T_e$, the two central absorption peaks merge in a single peak and the absorption exhibits a typical Mollow-like triplet profile~\cite{Ulhaq12}, as showed in the Fig.~\ref{fig:regimes}(d).

\begin{table}[ht]
\caption{Optical response behavior and hamiltonian parameter regime for the three optical regimes in a five-level QDM.}
\begin{tabular}{cc}
\hline
\hline
\textbf{Optical behavior} &\textbf{Parameter regime}\\
\hline
Quantum Interference (near TIT)&$T_e/\Gamma_4<1/4$\\
Intermediate (TIT-ATS)&$1/4<T_e/\Gamma_4<1$\\
Full ATS&$T_e/\Gamma_4>1$\\
Mollow-like Triplet&$T_e=\Omega_c$\\
\hline
\hline
\end{tabular}
\label{tab:threshold}
\end{table}
In conclusion, we have investigate in details the absorption of a QDM in the presence of a pump and probe laser fields. Our results show that the tunneling plays a crucial role in the optical response, creating different possibilities for quantum interference effects. In order to summarize our results, the parameters conditions for different optical behavior are listed in Table~\ref{tab:threshold}.  It becomes evident that the combined action of the external electric field, tunneling and coupling laser guarantees a high degree of control of the optical processes of the QDM. Our analysis established that a simple manipulation of the external fields allow us to control some quantum interference effects in such nanoestructure device. This has a potential application in optical devices, as for example, quantum memory.

This work was supported by the Brazilian National Institute of Science and Technology for Quantum Information (INCT-IQ) and for Semiconductor Nanodevices (INCT-DISSE), CAPES, FAPEMIG and CNPq.


\begin{thebibliography}{21}%
\makeatletter
\providecommand \@ifxundefined [1]{%
 \@ifx{#1\undefined}
}%
\providecommand \@ifnum [1]{%
 \ifnum #1\expandafter \@firstoftwo
 \else \expandafter \@secondoftwo
 \fi
}%
\providecommand \@ifx [1]{%
 \ifx #1\expandafter \@firstoftwo
 \else \expandafter \@secondoftwo
 \fi
}%
\providecommand \natexlab [1]{#1}%
\providecommand \enquote  [1]{``#1''}%
\providecommand \bibnamefont  [1]{#1}%
\providecommand \bibfnamefont [1]{#1}%
\providecommand \citenamefont [1]{#1}%
\providecommand \href@noop [0]{\@secondoftwo}%
\providecommand \href [0]{\begingroup \@sanitize@url \@href}%
\providecommand \@href[1]{\@@startlink{#1}\@@href}%
\providecommand \@@href[1]{\endgroup#1\@@endlink}%
\providecommand \@sanitize@url [0]{\catcode `\\12\catcode `\$12\catcode
  `\&12\catcode `\#12\catcode `\^12\catcode `\_12\catcode `\%12\relax}%
\providecommand \@@startlink[1]{}%
\providecommand \@@endlink[0]{}%
\providecommand \url  [0]{\begingroup\@sanitize@url \@url }%
\providecommand \@url [1]{\endgroup\@href {#1}{\urlprefix }}%
\providecommand \urlprefix  [0]{URL }%
\providecommand \Eprint [0]{\href }%
\providecommand \doibase [0]{http://dx.doi.org/}%
\providecommand \selectlanguage [0]{\@gobble}%
\providecommand \bibinfo  [0]{\@secondoftwo}%
\providecommand \bibfield  [0]{\@secondoftwo}%
\providecommand \translation [1]{[#1]}%
\providecommand \BibitemOpen [0]{}%
\providecommand \bibitemStop [0]{}%
\providecommand \bibitemNoStop [0]{.\EOS\space}%
\providecommand \EOS [0]{\spacefactor3000\relax}%
\providecommand \BibitemShut  [1]{\csname bibitem#1\endcsname}%
\let\auto@bib@innerbib\@empty
%</preamble>
\bibitem [{\citenamefont {Yang}\ \emph {et~al.}(2005)\citenamefont {Yang},
  \citenamefont {Zhang}, \citenamefont {Li}, \citenamefont {Han}, \citenamefont
  {Fu}, \citenamefont {Manson}, \citenamefont {Suter},\ and\ \citenamefont
  {Wei}}]{Yang05}%
  \BibitemOpen
  \bibfield  {author} {\bibinfo {author} {\bibfnamefont {L.}~\bibnamefont
  {Yang}}, \bibinfo {author} {\bibfnamefont {L.}~\bibnamefont {Zhang}},
  \bibinfo {author} {\bibfnamefont {X.}~\bibnamefont {Li}}, \bibinfo {author}
  {\bibfnamefont {L.}~\bibnamefont {Han}}, \bibinfo {author} {\bibfnamefont
  {G.}~\bibnamefont {Fu}}, \bibinfo {author} {\bibfnamefont {N.~B.}\
  \bibnamefont {Manson}}, \bibinfo {author} {\bibfnamefont {D.}~\bibnamefont
  {Suter}}, \ and\ \bibinfo {author} {\bibfnamefont {C.}~\bibnamefont {Wei}},\
  }\href@noop {} {\bibfield  {journal} {\bibinfo  {journal} {Phys. Rev. A}\
  }\textbf {\bibinfo {volume} {72}},\ \bibinfo {pages} {053801} (\bibinfo
  {year} {2005})}\BibitemShut {NoStop}%
\bibitem [{\citenamefont {Mollow}(1969)}]{Mollow69}%
  \BibitemOpen
  \bibfield  {author} {\bibinfo {author} {\bibfnamefont {B.~R.}\ \bibnamefont
  {Mollow}},\ }\href@noop {} {\bibfield  {journal} {\bibinfo  {journal} {Phys.
  Rev.}\ }\textbf {\bibinfo {volume} {188}},\ \bibinfo {pages} {1969-1975} (\bibinfo
  {year} {1969})}\BibitemShut {NoStop}%
\bibitem [{\citenamefont {Ulhaq}\ \emph {et~al.}(2012)\citenamefont {Ulhaq},
  \citenamefont {Weiler}, \citenamefont {Ulrich}, \citenamefont
  {R.Ro{\ss}bach}, \citenamefont {Jetter},\ and\ \citenamefont
  {Michler}}]{Ulhaq12}%
  \BibitemOpen
  \bibfield  {author} {\bibinfo {author} {\bibfnamefont {A.}~\bibnamefont
  {Ulhaq}}, \bibinfo {author} {\bibfnamefont {S.}~\bibnamefont {Weiler}},
  \bibinfo {author} {\bibfnamefont {S.~M.}\ \bibnamefont {Ulrich}}, \bibinfo
  {author} {\bibnamefont {R.Ro{\ss}bach}}, \bibinfo {author} {\bibfnamefont
  {M.}~\bibnamefont {Jetter}}, \ and\ \bibinfo {author} {\bibfnamefont
  {P.}~\bibnamefont {Michler}},\ }\href@noop {} {\bibfield  {journal} {\bibinfo
   {journal} {Nature Photonics}\ }\textbf {\bibinfo {volume} {6}},\ \bibinfo
  {pages} {238} (\bibinfo {year} {2012})}\BibitemShut {NoStop}%
\bibitem [{\citenamefont {Fleischhauer}, \citenamefont {Imamoglu},\ and\
  \citenamefont {Marangos}(2005)}]{Marangos05}%
  \BibitemOpen
  \bibfield  {author} {\bibinfo {author} {\bibfnamefont {M.}~\bibnamefont
  {Fleischhauer}}, \bibinfo {author} {\bibfnamefont {A.}~\bibnamefont
  {Imamoglu}}, \ and\ \bibinfo {author} {\bibfnamefont {J.~P.}\ \bibnamefont
  {Marangos}},\ }\href@noop {} {\bibfield  {journal} {\bibinfo  {journal}
  {Rev.\ Mod.\ Phys.}\ }\textbf {\bibinfo {volume} {77}},\ \bibinfo {pages}
  {633} (\bibinfo {year} {2005})}\BibitemShut {NoStop}%
\bibitem [{\citenamefont {Lu}, \citenamefont {Burkett},\ and\ \citenamefont
  {Xiao}(1998)}]{Lu98}%
  \BibitemOpen
  \bibfield  {author} {\bibinfo {author} {\bibfnamefont {B.}~\bibnamefont
  {Lu}}, \bibinfo {author} {\bibfnamefont {W.~H.}\ \bibnamefont {Burkett}}, \
  and\ \bibinfo {author} {\bibfnamefont {M.}~\bibnamefont {Xiao}},\ }\href@noop
  {} {\bibfield  {journal} {\bibinfo  {journal} {Optics Letters}\ }\textbf
  {\bibinfo {volume} {23}}\ \bibinfo {pages} {804-806} (\bibinfo {year} {1998})}\BibitemShut {NoStop}%
\bibitem [{\citenamefont {Ottaviani}\ \emph {et~al.}(2006)\citenamefont
  {Ottaviani}, \citenamefont {Rebic}, \citenamefont {Vitali},\ and\
  \citenamefont {Tombesi}}]{Ottaviani06}%
  \BibitemOpen
  \bibfield  {author} {\bibinfo {author} {\bibfnamefont {C.}~\bibnamefont
  {Ottaviani}}, \bibinfo {author} {\bibfnamefont {S.}~\bibnamefont {Rebic}},
  \bibinfo {author} {\bibfnamefont {D.}~\bibnamefont {Vitali}}, \ and\ \bibinfo
  {author} {\bibfnamefont {P.}~\bibnamefont {Tombesi}},\ }\href@noop {}
  {\bibfield  {journal} {\bibinfo  {journal} {Phys. Rev. A}\ }\textbf {\bibinfo
  {volume} {73}},\ \bibinfo {pages} {0103301} (\bibinfo {year}
  {2006})}\BibitemShut {NoStop}%
\bibitem [{\citenamefont {Hemmer}\ \emph {et~al.}(1995)\citenamefont {Hemmer},
  \citenamefont {Katz}, \citenamefont {Donoghue}, \citenamefont
  {Cronin-Golomb}, \citenamefont {Shahriar},\ and\ \citenamefont
  {Kumar}}]{Katz95}%
  \BibitemOpen
  \bibfield  {author} {\bibinfo {author} {\bibfnamefont {P.~R.}\ \bibnamefont
  {Hemmer}}, \bibinfo {author} {\bibfnamefont {D.~P.}\ \bibnamefont {Katz}},
  \bibinfo {author} {\bibfnamefont {J.}~\bibnamefont {Donoghue}}, \bibinfo
  {author} {\bibfnamefont {M.}~\bibnamefont {Cronin-Golomb}}, \bibinfo {author}
  {\bibfnamefont {M.~S.}\ \bibnamefont {Shahriar}}, \ and\ \bibinfo {author}
  {\bibfnamefont {P.}~\bibnamefont {Kumar}},\ }\href@noop {} {\bibfield
  {journal} {\bibinfo  {journal} {Optics Letters}\ }\textbf {\bibinfo {volume}
  {20}}\ \bibinfo {pages} {982-984} (\bibinfo {year} {1995})}\BibitemShut {NoStop}%
\bibitem [{\citenamefont {Wang}\ \emph {et~al.}(2004)\citenamefont {Wang},
  \citenamefont {Kong}, \citenamefont {Tu}, \citenamefont {Jiang},
  \citenamefont {Li}, \citenamefont {Xiong}, \citenamefont {Zhu},\ and\
  \citenamefont {Zhan}}]{Wang04}%
  \BibitemOpen
  \bibfield  {author} {\bibinfo {author} {\bibfnamefont {J.}~\bibnamefont
  {Wang}}, \bibinfo {author} {\bibfnamefont {L.~B.}\ \bibnamefont {Kong}},
  \bibinfo {author} {\bibfnamefont {X.~H.}\ \bibnamefont {Tu}}, \bibinfo
  {author} {\bibfnamefont {K.~J.}\ \bibnamefont {Jiang}}, \bibinfo {author}
  {\bibfnamefont {K.}~\bibnamefont {Li}}, \bibinfo {author} {\bibfnamefont
  {H.~W.}\ \bibnamefont {Xiong}}, \bibinfo {author} {\bibfnamefont
  {Y.}~\bibnamefont {Zhu}}, \ and\ \bibinfo {author} {\bibfnamefont {M.~S.}\
  \bibnamefont {Zhan}},\ }\href@noop {} {\bibfield  {journal} {\bibinfo
  {journal} {Phys. Lett. A}\ }\textbf {\bibinfo {volume} {328}},\ \bibinfo
  {pages} {437} (\bibinfo {year} {2004})}\BibitemShut {NoStop}%
\bibitem [{\citenamefont {Doty}\ \emph {et~al.}(2008)\citenamefont {Doty},
  \citenamefont {Scheibner}, \citenamefont {Bracker}, \citenamefont
  {Ponomarev}, \citenamefont {Reinecke},\ and\ \citenamefont
  {Gammon}}]{Doty08}%
  \BibitemOpen
  \bibfield  {author} {\bibinfo {author} {\bibfnamefont {M.~F.}\ \bibnamefont
  {Doty}}, \bibinfo {author} {\bibfnamefont {M.}~\bibnamefont {Scheibner}},
  \bibinfo {author} {\bibfnamefont {A.~S.}\ \bibnamefont {Bracker}}, \bibinfo
  {author} {\bibfnamefont {I.~V.}\ \bibnamefont {Ponomarev}}, \bibinfo {author}
  {\bibfnamefont {T.~L.}\ \bibnamefont {Reinecke}}, \ and\ \bibinfo {author}
  {\bibfnamefont {D.}~\bibnamefont {Gammon}},\ }\href@noop {} {\bibfield
  {journal} {\bibinfo  {journal} {Phys. Rev. B}\ }\textbf {\bibinfo {volume}
  {78}},\ \bibinfo {pages} {115316} (\bibinfo {year} {2008})}\BibitemShut
  {NoStop}%
\bibitem [{\citenamefont {Borges}\ \emph {et~al.}(2012)\citenamefont {Borges},
  \citenamefont {Sanz}, \citenamefont {Villas-Bo\^as},\ and\ \citenamefont
  {Alcalde}}]{Borges12}%
  \BibitemOpen
  \bibfield  {author} {\bibinfo {author} {\bibfnamefont {H.~S.}\ \bibnamefont
  {Borges}}, \bibinfo {author} {\bibfnamefont {L.}~\bibnamefont {Sanz}},
  \bibinfo {author} {\bibfnamefont {J.~M.}\ \bibnamefont {Villas-Bo\^as}}, \
  and\ \bibinfo {author} {\bibfnamefont {A.~M.}\ \bibnamefont {Alcalde}},\
  }\href@noop {} {\bibfield  {journal} {\bibinfo  {journal} {Phys. Rev. B}\
  }\textbf {\bibinfo {volume} {85}},\ \bibinfo {pages} {115425} (\bibinfo
  {year} {2012})}\BibitemShut {NoStop}%
\bibitem [{\citenamefont {Bracker}\ \emph {et~al.}(2006)\citenamefont
  {Bracker}, \citenamefont {Scheibner}, \citenamefont {Doty}, \citenamefont
  {Stinaff}, \citenamefont {Ponomarev}, \citenamefont {Kim}, \citenamefont
  {Whitman}, \citenamefont {Reinecke},\ and\ \citenamefont
  {Gammon}}]{Bracker06}%
  \BibitemOpen
  \bibfield  {author} {\bibinfo {author} {\bibfnamefont {A.~S.}\ \bibnamefont
  {Bracker}}, \bibinfo {author} {\bibfnamefont {M.}~\bibnamefont {Scheibner}},
  \bibinfo {author} {\bibfnamefont {M.~F.}\ \bibnamefont {Doty}}, \bibinfo
  {author} {\bibfnamefont {E.~A.}\ \bibnamefont {Stinaff}}, \bibinfo {author}
  {\bibfnamefont {I.~V.}\ \bibnamefont {Ponomarev}}, \bibinfo {author}
  {\bibfnamefont {J.~C.}\ \bibnamefont {Kim}}, \bibinfo {author} {\bibfnamefont
  {L.~J.}\ \bibnamefont {Whitman}}, \bibinfo {author} {\bibfnamefont {T.~L.}\
  \bibnamefont {Reinecke}}, \ and\ \bibinfo {author} {\bibfnamefont
  {D.}~\bibnamefont {Gammon}},\ }\href@noop {} {\bibfield  {journal} {\bibinfo
  {journal} {Appl. Phys. Lett.}\ }\textbf {\bibinfo {volume} {89}},\ \bibinfo
  {pages} {233110} (\bibinfo {year} {2006})}\BibitemShut {NoStop}%
\bibitem [{\citenamefont {P.~Borri}\ \emph {et~al.}(2003)\citenamefont
  {P.~Borri}, \citenamefont {Woggon}, \citenamefont {Schwab},\ and\
  \citenamefont {Bayer}}]{Borri03}%
  \BibitemOpen
  \bibfield  {author} {\bibinfo {author} {\bibfnamefont {W.~L.}\ \bibnamefont
  {P.~Borri}}, \bibinfo {author} {\bibfnamefont {U.}~\bibnamefont {Woggon}},
  \bibinfo {author} {\bibfnamefont {M.}~\bibnamefont {Schwab}}, \ and\ \bibinfo
  {author} {\bibfnamefont {M.}~\bibnamefont {Bayer}},\ }\href@noop {}
  {\bibfield  {journal} {\bibinfo  {journal} {Phys. Rev. Lett.}\ }\textbf
  {\bibinfo {volume} {91}},\ \bibinfo {pages} {267401} (\bibinfo {year}
  {2003})}\BibitemShut {NoStop}%
\bibitem [{\citenamefont {Bardot}\ \emph {et~al.}(2005)\citenamefont {Bardot},
  \citenamefont {Schwab}, \citenamefont {Bayer}, \citenamefont {Fafard},
  \citenamefont {Wasilewski},\ and\ \citenamefont {Hawrylak}}]{Bardot05}%
  \BibitemOpen
  \bibfield  {author} {\bibinfo {author} {\bibfnamefont {C.}~\bibnamefont
  {Bardot}}, \bibinfo {author} {\bibfnamefont {M.}~\bibnamefont {Schwab}},
  \bibinfo {author} {\bibfnamefont {M.}~\bibnamefont {Bayer}}, \bibinfo
  {author} {\bibfnamefont {S.}~\bibnamefont {Fafard}}, \bibinfo {author}
  {\bibfnamefont {Z.}~\bibnamefont {Wasilewski}}, \ and\ \bibinfo {author}
  {\bibfnamefont {P.}~\bibnamefont {Hawrylak}},\ }\href@noop {} {\bibfield
  {journal} {\bibinfo  {journal} {Phys. Rev. B}\ }\textbf {\bibinfo {volume}
  {72}},\ \bibinfo {pages} {035314} (\bibinfo {year} {2005})}\BibitemShut
  {NoStop}%
\bibitem [{\citenamefont {Butov}\ \emph {et~al.}(1994)\citenamefont {Butov},
  \citenamefont {Zrenner}, \citenamefont {Abstreiter}, \citenamefont {Bohm},\
  and\ \citenamefont {Weimann}}]{Butov94}%
  \BibitemOpen
  \bibfield  {author} {\bibinfo {author} {\bibfnamefont {L.~V.}\ \bibnamefont
  {Butov}}, \bibinfo {author} {\bibfnamefont {A.}~\bibnamefont {Zrenner}},
  \bibinfo {author} {\bibfnamefont {G.}~\bibnamefont {Abstreiter}}, \bibinfo
  {author} {\bibfnamefont {G.}~\bibnamefont {Bohm}}, \ and\ \bibinfo {author}
  {\bibfnamefont {G.}~\bibnamefont {Weimann}},\ }\href@noop {} {\bibfield
  {journal} {\bibinfo  {journal} {Phys. Rev. Lett.}\ }\textbf {\bibinfo
  {volume} {73}},\ \bibinfo {pages} {304} (\bibinfo {year} {1994})}\BibitemShut
  {NoStop}%
\bibitem [{\citenamefont {Rolon}\ and\ \citenamefont {Ulloa}(2010)}]{Rolon10}%
  \BibitemOpen
  \bibfield  {author} {\bibinfo {author} {\bibfnamefont {J.~E.}\ \bibnamefont
  {Rolon}}\ and\ \bibinfo {author} {\bibfnamefont {S.~E.}\ \bibnamefont
  {Ulloa}},\ }\href@noop {} {\bibfield  {journal} {\bibinfo  {journal} {Phys.
  Rev. B}\ }\textbf {\bibinfo {volume} {82}},\ \bibinfo {pages} {1153070}
  (\bibinfo {year} {2010})}\BibitemShut {NoStop}%
\bibitem [{\citenamefont {Tackeuchi}\ \emph {et~al.}(2000)\citenamefont
  {Tackeuchi}, \citenamefont {Kuroda}, \citenamefont {Mase}, \citenamefont
  {Nakata},\ and\ \citenamefont {Yokoyama}}]{Nakata00}%
  \BibitemOpen
  \bibfield  {author} {\bibinfo {author} {\bibfnamefont {A.}~\bibnamefont
  {Tackeuchi}}, \bibinfo {author} {\bibfnamefont {T.}~\bibnamefont {Kuroda}},
  \bibinfo {author} {\bibfnamefont {K.}~\bibnamefont {Mase}}, \bibinfo {author}
  {\bibfnamefont {Y.}~\bibnamefont {Nakata}}, \ and\ \bibinfo {author}
  {\bibfnamefont {N.}~\bibnamefont {Yokoyama}},\ }\href@noop {} {\bibfield
  {journal} {\bibinfo  {journal} {Phys. Rev. B}\ }\textbf {\bibinfo {volume}
  {62}},\ \bibinfo {pages} {1568} (\bibinfo {year} {2000})}\BibitemShut
  {NoStop}%
\bibitem [{\citenamefont {Emary}\ and\ \citenamefont {Sham}(2007)}]{Emay07}%
  \BibitemOpen
  \bibfield  {author} {\bibinfo {author} {\bibfnamefont {C.}~\bibnamefont
  {Emary}}\ and\ \bibinfo {author} {\bibfnamefont {L.~J.}\ \bibnamefont
  {Sham}},\ }\href@noop {} {\bibfield  {journal} {\bibinfo  {journal} {Phys.
  Rev. B}\ }\textbf {\bibinfo {volume} {75}},\ \bibinfo {pages} {125317}
  (\bibinfo {year} {2007})}\BibitemShut {NoStop}%
\bibitem [{\citenamefont {Kim}\ \emph {et~al.}(2004)\citenamefont {Kim},
  \citenamefont {Chuang}, \citenamefont {Ku},\ and\ \citenamefont
  {Chang-Hasnain}}]{Kim04}%
  \BibitemOpen
  \bibfield  {author} {\bibinfo {author} {\bibfnamefont {J.}~\bibnamefont
  {Kim}}, \bibinfo {author} {\bibfnamefont {S.~L.}\ \bibnamefont {Chuang}},
  \bibinfo {author} {\bibfnamefont {P.~C.}\ \bibnamefont {Ku}}, \ and\ \bibinfo
  {author} {\bibfnamefont {C.~J.}\ \bibnamefont {Chang-Hasnain}},\ }\href@noop
  {} {\bibfield  {journal} {\bibinfo  {journal} {J. Phys: Condens. Matter}\
  }\textbf {\bibinfo {volume} {16}},\ \bibinfo {pages} {S3727} (\bibinfo {year}
  {2004})}\BibitemShut {NoStop}%
\bibitem [{\citenamefont {Borges}\ \emph {et~al.}(2010)\citenamefont {Borges},
  \citenamefont {Sanz}, \citenamefont {Villas-B\^oas},\ and\ \citenamefont
  {Alcalde}}]{Borges10}%
  \BibitemOpen
  \bibfield  {author} {\bibinfo {author} {\bibfnamefont {H.~S.}\ \bibnamefont
  {Borges}}, \bibinfo {author} {\bibfnamefont {L.}~\bibnamefont {Sanz}},
  \bibinfo {author} {\bibfnamefont {J.~M.}\ \bibnamefont {Villas-B\^oas}}, \
  and\ \bibinfo {author} {\bibfnamefont {A.~M.}\ \bibnamefont {Alcalde}},\
  }\href {\doibase 10.1103/PhysRevB.81.075322} {\bibfield  {journal} {\bibinfo
  {journal} {Phys. Rev. B}\ }\textbf {\bibinfo {volume} {81}},\ \bibinfo
  {pages} {075322} (\bibinfo {year} {2010})}\BibitemShut {NoStop}%
\bibitem [{\citenamefont {Abi-salloum}\ \emph {et~al.}(2007)\citenamefont
  {Abi-salloum}, \citenamefont {Davis}, \citenamefont {Lehman}, \citenamefont
  {Elliott},\ and\ \citenamefont {Narducci}}]{Salloum07}%
  \BibitemOpen
  \bibfield  {author} {\bibinfo {author} {\bibfnamefont {T.}~\bibnamefont
  {Abi-salloum}}, \bibinfo {author} {\bibfnamefont {J.~P.}\ \bibnamefont
  {Davis}}, \bibinfo {author} {\bibfnamefont {C.}~\bibnamefont {Lehman}},
  \bibinfo {author} {\bibfnamefont {E.}~\bibnamefont {Elliott}}, \ and\
  \bibinfo {author} {\bibfnamefont {F.~A.}\ \bibnamefont {Narducci}},\ }\href
  {\doibase 10.1080/09500340701742617} {\bibfield  {journal} {\bibinfo
  {journal} {Journal of Modern Optics}\ }\textbf {\bibinfo {volume} {54}},\
  \bibinfo {pages} {2459} (\bibinfo {year} {2007})}\BibitemShut {NoStop}%
\bibitem [{\citenamefont {Abi-Salloum}(2010)}]{Salloum10}%
  \BibitemOpen
  \bibfield  {author} {\bibinfo {author} {\bibfnamefont {T.~Y.}\ \bibnamefont
  {Abi-Salloum}},\ }\href@noop {} {\bibfield  {journal} {\bibinfo  {journal}
  {Phys. Rev. A}\ }\textbf {\bibinfo {volume} {81}},\ \bibinfo {pages} {053836}
  (\bibinfo {year} {2010})}\BibitemShut {NoStop}%
\end{thebibliography}
\end{document}